\documentclass[conference]{IEEEtran}
\usepackage{cite}
\usepackage{comment}
\usepackage{graphicx}
\usepackage{caption}
\usepackage{subcaption}
\usepackage{caption}
\usepackage{amsthm}
\usepackage[table,dvipsnames]{xcolor}
\usepackage{amssymb}
\usepackage{mathrsfs}
\usepackage{mathtools}

\usepackage[normalem]{ulem}
\usepackage{cancel}

\usepackage{hhline}
\usepackage{tabularx,colortbl}

\usepackage{comment}

\usepackage{multirow}
\usepackage{booktabs}
\usepackage[noend]{algpseudocode}
\usepackage{algorithm}

\newtheorem{thm}{Theorem}

\newtheorem{defin}{Definition}
\newtheorem{assumption}{Assumption}

\def\CC{\mathbb C}
\def\RR{\mathbb R}

\def\cA{\mathcal A}

\def\cD{\mathcal D}

\def\cG{\mathcal G}

\def\cM{\mathcal M}
\def\cN{\mathcal N}

\def\cT{\mathcal T}

\newcommand{\hide}[1]{}
\newcommand{\raf}[1]{(\ref{#1})}

\newcommand{\re}{\ensuremath{\mathrm{Re}}}
\newcommand{\im}{\ensuremath{\mathrm{Im}}}

\newcommand{\cV}{\ensuremath{\mathcal{V}}}
\newcommand{\cE}{\ensuremath{\mathcal{E}} }

\newcommand{\Opt}{\ensuremath{\textsc{Opt}}}

\DeclareCaptionType{copyrightbox}

\title{Assessing the Privacy Cost in Centralized Event-Based Demand Response for Microgrids }

\author{
	\IEEEauthorblockN{Areg Karapetyan, Syafiq Kamarul Azman, and Zeyar Aung}
	\IEEEauthorblockA{Deparment of Electrical Engineering and Computer Science}
	\IEEEauthorblockA{Masdar Institute of Science and Technology}
	\IEEEauthorblockA{Abu Dhabi, United Arab Emirates}
		emails:\{akarapetyan, mbinkamarulazman, zaung\}@masdar.ac.ae
	
}
\begin{document}
\maketitle

\begin{abstract}
Demand response (DR) programs have emerged as a potential key enabling ingredient in the context of smart grid (SG). Nevertheless, the rising concerns over privacy issues raised by customers subscribed to these programs constitute a major threat towards their effective deployment and utilization. This has driven extensive research to resolve the hindrance confronted, resulting in a number of methods being proposed for preserving customers' privacy. While these methods provide stringent privacy guarantees, only limited attention has been paid to their computational efficiency and performance quality. Under the paradigm of differential privacy, this paper initiates a systematic empirical study on quantifying the trade-off between privacy and optimality in centralized DR systems for maximizing cumulative customer utility. Aiming to elucidate the factors governing this trade-off, we analyze the cost of privacy in terms of the effect incurred on the objective value of the DR optimization problem when applying the employed privacy-preserving strategy based on Laplace mechanism. The theoretical results derived from the analysis are complemented with empirical findings, corroborated extensively by simulations on a $4$-bus MG system with up to thousands of customers. By evaluating the impact of privacy, this pilot study serves DR practitioners when considering the social and economic implications of deploying privacy-preserving DR programs in practice. Moreover, it stimulates further research on exploring more efficient approaches with bounded performance guarantees for optimizing energy procurement of MGs without infringing the privacy of customers on demand side.
\end{abstract}


\section{Introduction}\label{intro}

Rising environmental and economic concerns necessitate modernization of the aging power grid infrastructure into a more sustainable and optimized cyber-physical system, SG. Microgrids, reckoned as a vital contributor towards this transition, facilitate large scale deployment of renewable Distributed Generation (DG) and incorporation of new load types such as plug-in hybrid electric vehicles (PHEVs). However, the volatile nature of renewable energy sources along with increasing customer expectation for both power quality and quantity further complicate the energy management of MGs besetting with an intricate power allocation problem critical for maintaining system stability.

DR management has proven instrumental in resolving the problem confronted, offering efficient schemes for energy procurement and optimization of MGs suffering from power disbalance \cite{5454394}. To achieve the desired system reliability, resilience and power quality, DR programs aim at establishing a mutually beneficial interaction framework for DR participants and aggregators where power generation drives the demand. With DR, customers are incentivized to shape and schedule their consumption profiles to flatten the peak demand, consequently deferring the cost of generation expansion and ancillary grid services. Among other advantages, DR enables reduced price variations \cite{Ref1}, enhanced congestion management \cite{Ref3} and strengthened system security \cite{Ref4}. DR programs can be broadly categorized into two classes, price-based and event-based\cite{Siano2014461}. Event-based DR programs are efficient especially during the times when there is insufficient supply of power on MG side to meet the available demand\cite{6574273, Siano2014461}.

Increasing customer participation translates into elevated benefits for DR operators, as the availability and volume of controllable load reserves grows with the number of DR participant devices. To this end, customers' concerns over privacy could constitute a major threat\cite{hoenkamp2011neglected, Molina}. DR programs with centralized control architecture, which are commonly deployed in practice \cite{5759191, 6861959}, entail information pertaining to customer loads and preferences as the inputs for optimization. This may limit customer participation severely since customers may not be willing to reveal such information out of confidentiality reasons\cite{7069275}. Possessing knowledge on the desired power valuation (utility) and electricity consumption characteristics of a customer empowers DR operators or potential eavesdroppers to exploit this marketable information to their own benefit. According to\cite{6847974}, information on power consumption alone suffices to infer customer's appliances operated or even daily activities. More importantly, with this information an adversary participant may attempt to manipulate the DR outcome by misreporting own utility value for the sake of profit.  

It is widely acknowledged that ad-hoc approaches such as anonymizing customer data are devoid of privacy features owing to the availability of public side information\cite{Narayanan:2008}. Alternatively, it might be tempting to retain privacy in DR with a distributed control scheme in place, where response decisions are computed on a customer's end involving exchange of surrogate information only. In\cite{6840288} a distributed energy management strategy is proposed with the objective of maximizing aggregate utilities of the appliances in the households and minimizing power losses. However, distributed DR programs typically suffer from synchronization problems and could incur diminished system-wide controllability and efficiency, in contrast to the centralized ones\cite{5759191}. Besides, it is suspected that, in general, inferring the private customer information concealed implicitly in the exchanged surrogate coordination signals could be still viable. In the view of the above arguments, deriving efficient privacy-aware strategies for solving large-scale DR problems under centralized control philosophy becomes vital.

While ensuring customer privacy, DR operators should also seek to achieve certain benefits determined by the objective function of DR optimization problem. As long as the private information lies in the objective function or constraint matrix of the DR problem these two objectives are conflicting. Indeed, optimizing the DR management choosing to ignore customer input data as a privacy-preventive measure may lead to arbitrarily worse solutions when compared to the optimal solution. The induced suboptimality gap typifies the cost of privacy, in a sense to be formalized in Section~\ref{alg}. Most of the extant literature on energy management of MGs, such as\cite{7069275, 5951960, gong2016privacy} and\cite{6175785, 4454000, 1626398}, approach this trade-off from one angle or the other leaving the privacy cost largely unexamined. Among these works, those dealing with the privacy aspect advocate approaches relying on cryptographic techniques and security protocols that may evoke substantial communication and computation overhead, thereby questioning their practicality in large-scale applications. 

Recently, smart metering infrastructure sparked considerable research efforts\cite{Gulisano:2016, 5503916, 6514815, 6203629} in response to privacy threats. Various schemes have been developed for securing the smart meter data aggregation and customer demand reporting in SG. However, the setting in these studies envisions smart metering primarily in the scope of electricity billing service rather than deployed within a DR optimization model as in the case studied here. A simple yet effective privacy solutions are presented in\cite{kalogridis2010privacy, mclaughlin2011protecting} for demand reporting by utilizing rechargeable batteries. Essentially, the batteries are used as a proxy between smart metering devices and household appliances to mask consumer demands in a non-intrusive manner. This is desirable as it maintains consumer privacy without introducing additional noise while satisfying smart grid constraints.

Against this background, the focus here is to explore the interplay between privacy and optimality in centralized event-based DR management programs through extensive empirical analysis. In addition to this, the gap between suboptimal private and optimal non-private solutions is bounded theoretically. We further unravel the obscure relationship between privacy and DR parameters driving this trade-off, thus conducing to deeper understanding of the scales and dynamics of privacy phenomena in event-based energy management for MGs. 

%We also provide theoretical
%analysis on the suboptimality of these mechanisms
%and show the trade-offs between optimality and privacy.

As to other related literature, studies in \cite{barbosa2016technique,loucost} covered a more thorough investigation of the privacy cost. An efficient privacy-assuring metering scheme is proposed in \cite{barbosa2016technique}. Customer's privacy is achieved by adding a randomly generated number to the measurement sent from the smart meter to the power provider, making it a simple and low complexity approach. The trade-off between economic benefit and privacy is evaluated in terms of the error in the billed amount to SG operator. Compared to\cite{barbosa2016technique}, this paper provides a deeper insight of the privacy effect on MG performance when evaluating it within a comprehensive DR optimization framework with an accurate realistic MG model and privacy levels adaptable to customer preferences. Privacy preservation is also evident in economic load dispatch control problem for minimizing generation cost\cite{loucost}. Customer demands are altered by noise prior to reaching the control unit to enable the privacy. The increased generation cost attributed to privacy is assessed through small scale experiments based on a 5-bus power system with $200$ customers only and a simplified grid model that omits AC power flow equations. 
     
Typically, in a DR management scheme, there is a single load-serving entity (LSE) or an operator of MG, who coordinates the decisions of DR participants. There is a high probability that an MG once initiated will be short of power, consequently resulting in significant voltage and frequency deviations, and leading to its instability. The LSE then invokes the featured event-based DR program to ensure the endurance of an
MG by balancing out power generation and demand. Constrained by the net available apparent generation and power flow equations, LSE is required to make control decisions in real time as to maximize the total utility of satisfied customers without violating their privacy. To attain this in a scalable fashion, an efficient privacy-preserving mechanism introduced in \cite{hsu2014privately} is leveraged to this end, which also provides a definite theoretical guarantee on the level of privacy and optimality. The privacy cost is quantified by benchmarking the maximized utility of this mechanism with that in the omniscient case, where customer privacy is not protected.

The major contribution of this preliminary study is centered on two salient features that differentiate it from the surveyed literature. First, particular emphasis is
paid to establishing a realistic DR system with an accurate and reliable MG model capturing power flow and operational constraints (e.g., Kirchhoff’s law, voltage bounds, reactive power) associated with the underlying distribution network. The importance and necessity of this were acknowledged in\cite{6980137, 6489288, 6805674}, which highlight that using a simplified MG model may lead to infeasible load management decisions in practice and thus impairs the credibility of the results produced. Second, power valuation, which is a core design parameter of an event-based DR, is regarded as information private to an individual customer only. While various approaches were proposed for tackling privacy in demand reporting (e.g., the method in \cite{kalogridis2010privacy, mclaughlin2011protecting} relying on rechargeable batteries), the setting with private power valuations has not been studied properly. Taken together, these contributions illustrate and appraise the ramifications of deploying event-based DR programs in practice where privacy concerns are a priority.

As one demonstration, the proposed privacy-preserving DR scheme is applied to a $4$-bus feeder from Canadian distribution system (see Fig.~\ref{fig:system} in Section~\ref{model}). The results indicate that incorporating privacy in centralized event-based DR management may degrade the optimality of the produced load management decisions severely. Nevertheless, considering a realistic scenario with heterogeneous privacy levels varying among customers may smooth the impact of privacy. Also, it is inferred that the privacy cost is influenced by several DR parameters including customer type, number of customers and privacy level.

%Taken together, these advances show that the proposed
%algorithm represents a feasible method for implementing
%large-scale demand response.
%
% Our
% analysis and cost sharing algorithms will provide an important
% basis for practical implementation of demand reporting for
% EDC and other smart grid control applications, by allowing
% customers to acquire sufficient privacy protection on a fair cost
% basis. 

%To the best of our knowledge, this
%is the first paper considering a concrete model of smart grid
%operations and the tradeoff between its performance and the
%privacy of its consumers

%the toolbox of general algorithmic techniques for designing computationally
%efficient and differentially private algorithms
%
%. First, we propose an intelligent algorithm that
%can effectively optimize energy consumption. In addition to
%reducing peak energy usage, we
%
%Second, the findings are promising for the continued development
%of more intelligent electricity management in the
%residential sector

\section{Preliminaries}

This section briefly scopes preliminaries of the adopted privacy notion.

\subsection{Differential Privacy}

When it comes to quantifying the extent of privacy of a customer participating in DR program, this paper adheres to the notion of \textit{differential privacy}. Originated from the research in \cite{dinur2003revealing} and defined by \cite{Dwork2006}, differential privacy has evolved as a rigorous definition of privacy in computer science. As such, it can be interpreted as a guarantee that altering an individual record in the input set does not impact the distribution of the computation outcomes significantly. In other terms, customers' private information becomes indistinguishable in the output of a differentially private algorithm. This yields a strong privacy guarantee regardless of any auxiliary information the adversary may possess. 

On the other hand, it is increasingly hard to derive efficient algorithms meeting such a stringent privacy guarantee\cite{hsu2014privately}. As a consequence, many fundamental problems as those studied in \cite{Blum:2008, kasiviswanathan2011can} have private solutions, while lacking efficient algorithms. For a certain class of combinatorial optimization problems several approximation algorithms are devised in\cite{Gupta:2010} that retain differential privacy. \textit{Laplace mechanism} and \textit{exponential mechanism}, which by far are the most prevalent approaches practiced for differential privacy were introduced in \cite{Dwork2006} and \cite{mcsherry2007mechanism}, respectively. Unlike the former one, exponential mechanism runs in linear time of its output range, which might largely outweigh the number of customers, and thus is not suitable for the purposes of this study.  

The framework of differential privacy represents information private to a customer as a set $D$ referred to as \textit{database}. Let $\cD$ be the domain of all databases of interest. The basic idea underlying this framework is to draw an association between privacy and impact of an individual customer in the database. The impact, that is, changes that occur in the database when altering or deleting a customer's record is characterized by the concept of \textit{neighboring databases}. Define databases $D \in \cD$ and $D^{\prime} \in \cD$ to be neighboring if they are identical except a single record. Differential privacy is formalized by the definition below.
\begin{defin}[Differential privacy\cite{Dwork2006}]
	A randomized algorithm $\cA:\cD \rightarrow \mathbb{R}^{n}$ that maps databases to an output range $\mathcal{R}$ is $(\epsilon, \delta)$-differentially private if for every pair of neighboring databases $D \in \cD$, $D^{\prime} \in \cD$ and $\forall ~S \subseteq \mathcal{R}$
	\begin{equation}\label{difeq}
	Pr[\cA(D) \in S] \leq e^{\epsilon}\cdot Pr[\cA(D^{\prime}) \in S] + \delta \, ,
	\end{equation}
	where $\epsilon >0$ and $\delta \in [0, 1)$.
\end{defin}
Similarly, an algorithm $\cA$ is called $\epsilon$-differentially private, if $\delta = 0$. The level of privacy is characterized by the constant $\epsilon$. The smaller the $\epsilon$ the higher is privacy level.

\section{System Model and Assumptions}\label{model}

Towards defining the proposed DR optimization problem formally, this section starts by modeling the system and its components. The adopted DR model envisions a single LSE procuring the responses of customers' demands over a decision horizon $\cT\triangleq\{1,...,m\}$. The decision horizon $\cT$ is discretized into $m$ equal periods with a duration corresponding to the required time resolution granularity at which DR management decisions are to be produced. At each time slot $t \in \cT$, the net available generation capacity of MG is denoted by $C_t \in \RR_{+}$.

\subsection{Load Model} 

Consider a set of customers $\cN\triangleq\{1,...,n\}$ for a DR management scheme run by LSE. A customer $k\in \cN$ is associated with a complex-valued power demand $S_{k} \in \CC$ required for operating certain electric appliances at particular time instant. Denote by $S_{k}^{\rm R} \triangleq \re(S_{k})$ the\textit{ active power} demand of customer $k$, and by $S_{k}^{\rm I} \triangleq \im(S_{k})$ the \textit{reactive power} demand. For clarity of presentation, this paper assumes (via a rotation of power demand vectors) that $S_{k}^{\rm R} \ge 0$ and $S_{k}^{\rm I} \ge 0$ for $\forall k\in \cN$. 

To allow effective DR application, diverse customer appliances should be taken into account. The customers' demands here are categorized into two types according to their operation and energy consumption characteristics, \textit{elastic} (divisible) and \textit{inelastic} (indivisible). The demand of a customer possessing inelastic load can be either shed or fed completely. This models the electric appliances that can operate only under particular energy supply level (e.g., washing machine, vacuum cleaner). Different from the inelastic demands, an elastic load may be satisfied partially and adjusted to operate with different energy consumption levels (e.g., air conditioner, LED light).

%To best capture the gains from DR and to ensure customer
%satisfaction, diverse usage profiles and operational constraints
%of the devices must be taken into account. 
%Three salient types
%of devices are considered in this work.

%The customers' demands are categorized into two types according to their operation and energy consumption characteristics, \textit{elastic} (divisible) and \textit{inelastic} (indivisible). The demand of a customer possessing inelastic load can be either shed or fed completely over the specified time period. This models the electric appliances that can operate only under particular energy supply level (e.g., washing machine, vacuum cleaner). Different from the inelastic demands, an elastic load may be satisfied partially and adjusted to operate with different energy consumption levels (e.g., air conditioners, LED light bulbs).

\subsection{Modeling the Distribution Network}

To incorporate the power flow and voltage constraints into the DR optimization problem a model of the distribution network, resembling that of in\cite{7590153}, is established below. We shall confine our attention to radial (tree) distribution networks which are common in practice\cite{schneider2008modern}.

The distribution system is represented by a graph $\cG=(\cV,\cE)$, where each customer $k \in \cN$ is located at a given node except the root. The set of nodes $\cV$ denote the electric buses, whereas the set of edges $\cE$ denote the distribution lines. The nodes in $\cV$ are indexed by $\{0,1,...,|\cV|\}$, where the node $0$ denotes the generation source of MG. Let $V_i \in \CC$ denote the voltage of node $i \in \cV$. Define $I_{i,j}$ to be the current flowing through edge $e=(i,j) \in \cE$ and with a slight abuse of notation, $\widehat S_{i,j} \in \CC$ to be the transmitted power through that edge. Similarly, let $z_{i,j} \in \CC$ be the impedance of that edge. Denote by 
$v_i \triangleq |V_i|^2$ and $\ell_{i,j} \triangleq |I_{i,j}|^2$ the magnitude square of voltage and current, respectively. For each node $i\in\cV\backslash\{0\}$, there is a set of customers attached to $i$, denoted by $\cM_i$.

A power flow in a steady state can be characterized by a set of power flow equations. In radial networks (which include paths) the Branch Flow Model (BFM) proposed by\cite{19265}
can be used to model them. Assuming $v_0$ is given, the BFM is captured by the following set of equations for $\forall~(i,j) \in \cE$ 
\begin{align}
\ell_{i,j}& =  \frac{|\widehat S_{i,j}|^2}{v_i},   \label{eqnc}\\
v_{j}& =  v_i + |z_{i,j}|^2 \ell_{i,j} - 2 \re(z_{i,j}^\ast  \widehat S_{i,j}),  \label{eq:vj}\\
\widehat S_{i,j}& = \sum_{l:(j,l)\in \cE}\widehat S_{j,l} + \sum_{k \in \cM_{j}} S_kx_k + z_{i,j}\ell_{i,j},  \label{eq:bf3}
% S^{\rm G} &=- \sum_{j: (0,j)\in \cE}\widehat S_{0,j}
\end{align}
where $x_k$ is the load management decision produced by LSE for customer $k \in \cN$,  $\re(\psi)$ denotes the real component of a complex number $\psi \in \CC$ and $\psi^\ast$ denotes the complex conjugate of $\psi$. Decision variable $x_k$ takes values either from $[0,1]$ or $\{0, 1\}$ for elastic and inelastic demands, respectively. Equations~\raf{eqnc} - \raf{eq:bf3}, essentially, capture Ohm's law combined with the Kirchhoff’s laws of electric flows and power flow definitions.

\begin{figure}[!htb] \vspace{-5pt}
	\begin{center}
		\includegraphics[scale=.43]{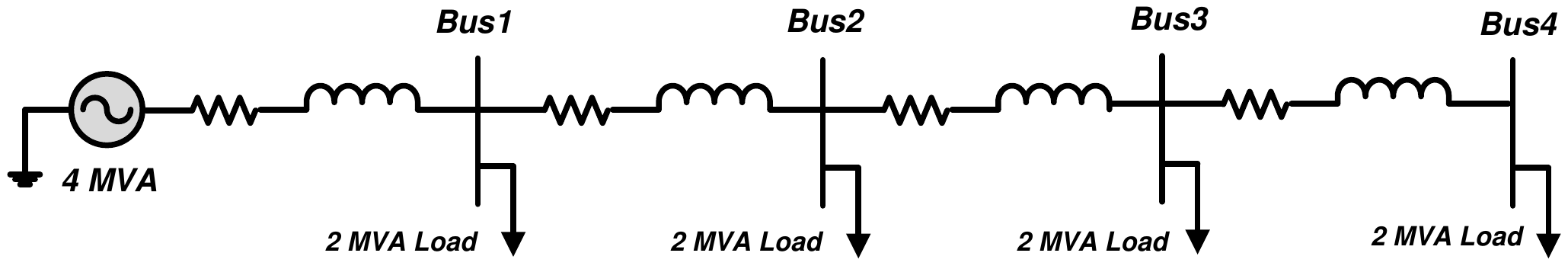}
	\end{center} \vspace{-10pt}
	\caption{A 4-bus feeder from Canadian benchmark distribution system.}
	\label{fig:system}
\end{figure}

Furthermore, for each node $i \in \cV \setminus \{0\}$ the following operational constrain should be satisfied
\begin{align}
v_{\min} \leq v_i \leq v_{\max} \label{eq4}\,,
\end{align}
where $v_{\min}, v_{\max} \in \mathbb{R}^{+}$ are the minimum and maximum allowable voltage magnitude square at any node, respectively. The setting studied here assumes a limited apparent power generation on MG and thus at each time step $t \in \cT$
\begin{align}
|\widehat S_{0,1}| \leq C_t\label{eq5}\,.
\end{align}
Observe that BFM model is non-convex due to the quadratic equality constraints in Eqn.\raf{eqnc} and thus is computationally intractable in general. We therefore consider relaxing them to inequalities in Eqn.~\raf{eqnewl} to convexify the model. Due to the same reason, this relaxation is adopted in~\cite{6980137,6756976, 6489288}. Obviously, when the equality in~\raf{eqnewl} is attained then the relaxation is exact. In fact, 
it was shown in\cite{6843918} that this relaxation happens to be exact under certain mild conditions in most of the practical radial distribution systems including IEEE bus systems.
\begin{eqnarray}
\ell_{i,j} \geq \frac{|\widehat{S}_{i,j}|^2}{v_i}\, ,  \qquad \forall (i,j) \in \cE \label{eqnewl}
\end{eqnarray}
Lastly, it is assumed that the MG system encompasses a hybrid mix of traditional and renewable energy (RE) supplies that could collectively have a variable (depending on the availability of RE and storage available) yet dispatchable capacity. In the sequel, using the established distribution model, the employed differentially private mechanism of \cite{hsu2014privately} is applied to a $4$-bus feeder of the Canadian benchmark system depicted in Fig.~\ref{fig:system}.
\subsection{Customer Preference Model} 

In a DR program where each subscribed customer is an independent decision maker, typically, the response to the incentives by the DR aggregator is modeled by an \textit{utility function}. The response may vary depending on the particular time of a day (e.g. at peak and non-peak hours). Moreover, it may vary from customer to customer based on the consumption profile (i.e., when considering \textit{residential} and \textit{commercial} customers).

To simplify the exposition, the utility function is summarized by an \textit{utility value} (power valuation) $u_{k}^t$ associated with a customer $k\in \cN$. This value quantifies the extent of satisfaction obtained (or alternatively, the payment) by customer $k$ when own power demand is satisfied at time $t \in \cT$. In the case with inelastic demands, if $S_{k}$ is satisfied at time slot $t$, $u_{k}^t$ is the perceived utility for customer $k$, otherwise zero utility is perceived. As for a customer $k^{\prime} \in \cN$ with an elastic load, a portion $b \in [0,1]$ of the power demand $S_{k^{\prime}}$ drawn from MG at time instant $t$ imparts an utility of $b\cdot u_{k^{\prime}}^t$.

\subsection{Privacy-Preserving DR}

Recall that the DR scheme under study requires protecting customer sensitive information in the input data. In particular, this paper focuses on the case where utilities, which define the objective function of DR optimization problem, are the sensitive information to customers. In more detail, the private database at time $t$ of the participating customers is $D=\{u_{k}^t\}_{k \in \cN}$. Since only aggregate power demands are required by LSE, it is assumed that the customer loads can be protected in a non-intrusive manner by a method akin to those proposed in \cite{kalogridis2010privacy, mclaughlin2011protecting} relying on rechargeable batteries. In a sense, shifting between neighboring databases affects only the objective function of the DR problem leaving the constraint matrix unchanged.

In the envisioned DR program featuring a centralized control scheme, each customer declares his reactive and active power demand through the equipped smart metering infrastructure to LSE upon request. In order to prevent customer utilities from being exposed to LSE or a potential eavesdropper the adopted strategy produces a perturbed utility value to obfuscate customer's true valuation. Section~\ref{alg} scopes a detailed explanation of this mechanism and provides provable guarantees on its privacy level.

It is worthy to note, that this scheme is general enough to handle customers with multiple utilities (independent of each other) for corresponding subsets of own demands.
This allows customers to prioritize their demands based on the importance and value without altering the proposed problem formulation. To effectively control DR management, low-latency communication infrastructure between the LSE and customers using separate power supply will be utilized in the proposed scheme. Such an infrastructure is enabled by the standards of SG communication protocols\cite{hossain2012smart}. Additionally, it is to be assumed that LSE has full control over the on/off or adjusting operations of its customers' demands.

%One of the salient properties and strengths of differential privacy framework is that provided privacy guarantees hold regardless of adversary's strength or ancillary information possessed. 
As mentioned previously, devising computationally efficient algorithms in differential privacy with provable optimality guarantees is substantially difficult. In fact, it was shown in~\cite{hsu2014privately} that if one makes no assumptions on the sensitivity of the private data it is impossible to devise such an algorithm with non-trivial optimality guarantees. This necessitates the need for introducing the simplifying assumption stated hereunder that will be followed throughout this paper.
\begin{assumption}
	There exist positive $u_{\max}$ and $u_{\min}$ known to LSE {\it apriori} such that for $\forall ~k \in \cN$, $u_{\min} \leq u_{k}^t \leq u_{\max} ~~\forall t \in \cT.$
	\label{assum}
\end{assumption}
We remark that Assumption~\ref{assum} naturally holds in DR systems, since usually $u_{\min}$ and $u_{\max}$ are determined by LSE. The subsequent section defines the DR optimization problem and formalizes the privacy cost.

% and presents an efficient differentially private algorithm with provable guarantees on its privacy level and optimality.

%\input{problem}
\section{Problem Formulation and Chosen Approach}\label{alg}

\subsection{Optimization Problem}
Now that the system model is established, the {\em utility maximizing demand response problem} ({\sc UMDR}) at time $t \in \cT$ can be formulated in an optimal power flow framework by the following \textit{quadratically constrained mixed integer programming} (QCMIP) problem. 
\begin{align}
\textsc{(UMDR)}\quad &\max_{\substack{x_k, v_i,\ell_{i,j}, \widehat{S}_{i,j} \;\;}} \sum_{k \in \cN} u_k^t x_k  \notag \\
\text{subject to} \quad& \raf{eq:vj}, \raf{eq:bf3}, \raf{eq4}, \raf{eq5}, \raf{eqnewl}\\
& x_k \in \{0,1\},\qquad \forall ~k \in \cN
\end{align}	
Here, $x_k$ is a binary decision variable that takes value $1$ if and only if the $k$-th customer's power demand $S_{k}$ is satisfied and $0$ otherwise. The {\sc UMDR} problem aims at maximizing the overall net utility of customers while maintaining the apparent power generation $C_t$ bound at time instant $t \in \cT$, power flow equations and voltage levels. 

Evidently, {\sc UMDR} is {\sc NP-hard}, since the $0$-$1$ classical knapsack problem is its special case. Relaxing binary (discrete)  decision variables ($x_k$) to continuous ones in {\sc UMDR} problem, such that $x_k \in [0, 1]$, yields a {\it convex quadratic programming} problem denoted by {\sc UMDR$_{\textsc{L}}$}. This seemingly small change in problem formulation, in fact, alters the complexity of DR optimization problem notably, since the convex variant can be solved optimally in polynomial time (e.g., by applying Interior Point methods). Aside from complexity, setting the decision variable ($x_k$) to be discrete or continuous alters the practical aspects of DR application. Concretely, the continuous case corresponds to customers having only elastic power demands. As for the case with binary decision variables, DR participant demands comprising the customer set are solely inelastic loads.

%Before the proposed privacy preserving mechanism is introduced, a performance metric for assessing the level of privacy of the computed solution of {\sc UMDR} and {\sc UMDR$_{\textsc{L}}$} problems is provided in the subsequent subsection.

\subsection{Differentially Private Method}
This subsection presents an efficient randomized mechanism, introduced by \cite{hsu2014privately}, to compute solutions of {\sc UMDR} and {\sc UMDR$_\textsc{L}$} problems privately. The method relies on a principle differential privacy technique which is explained in what follows.

Define a function $f:\cD\rightarrow \RR^n$ to be $\triangle$-sensitive if $||f(D) - f(D^{\prime})||_1 \leq \triangle \text{ for }\forall \text{ neighboring } D, D^{\prime} \in \cD$. Let $Lap(\Omega)$ denote the Laplace transformation of $\Omega$ with a probability density function $f(x\mid\Omega) = \frac{1}{2\Omega}e^{-\frac{|x|}{\Omega}}$. Then, applying the Laplace mechanism to a $\triangle$-sensitive function $f(D)$ yields $f(D) + [\mu_1,\mu_2,...,\mu_n]^T$, where $\mu_1,\mu_2,...,\mu_n$ are $n$ independent and identically distributed draws from $Lap(\frac{\triangle}{\epsilon})$ with $\epsilon > 0$.
\begin{thm}[\cite{Dwork2008}] A randomized algorithm $\cA$ that invokes the Laplace mechanism explained above is $\epsilon$-differentially private.
\end{thm}
Instead of releasing the true customer valuations, the considered mechanism perturbs each customer's utility $u_{k}^t$ independently by adding a noise drawn from the Laplace distribution that commensurates with the desired level of privacy. Define $\hat{u}_{k}^t \triangleq u_{k}^t + \text{Lap}(\frac{(u_{\max}-u_{\min})\sqrt{8n\log(\frac{1}{\delta})}}{\epsilon})$ to be the perturbed utility of customer $k \in \cN$ at time $t \in \cT$. Then the private analog of {\sc UMDR$_\textsc{L}$} problem (DR with elastic demands) is embodied by the following \textit{convex programming problem}.
\begin{align}
\textsc{(UMPDR)}_\textsc{L}\quad &\max_{\substack{x_k, v_i,\ell_{i,j}, \widehat{S}_{i,j} \;\;}} \sum_{k \in \cN} u_k^t x_k  \notag \\
\text{subject to} \quad& \raf{eq:vj}, \raf{eq:bf3}, \raf{eq4}, \raf{eq5}, \raf{eqnewl}\\
& x_k \in [0,1],\qquad \forall ~k \in \cN
\end{align}
Solving {\sc UMPDR$_\textsc{L}$} problem non-privately results in a private solution to the original one. Note that any feasible solution to {\sc UMPDR$_\textsc{L}$} is also feasible for {\sc UMDR$_\textsc{L}$}. Denote by $X_\textsc{L}^\ast \subseteq \cN$ an optimal solution of {\sc UMDR$_\textsc{L}$} and by $\Opt_{\textsc{L}} \triangleq \sum_{k \in X_\textsc{L}^\ast} u_k^t$ the corresponding total utility for any time $t \in \cT$.
Set $\hat{X}_\textsc{L}^\ast \subseteq N$ to be an optimal solution of {\sc UMPDR$_\textsc{L}$} and define $\Opt_{\textsc{L}}^{\textsc{DP}} \triangleq \sum_{k \in \hat{X}_\textsc{L}^\ast} u_k^t$.
\begin{thm}[\cite{hsu2014privately}]\label{thm} An optimal solution $\hat{X}_\textsc{L}^\ast$ for {\sc UMPDR$_\textsc{L}$} problem is $(\epsilon, \delta)$-differentially private feasible solution to {\sc UMDR$_\textsc{L}$} that with high probability satisfies the following additive optimality bound $\Opt_{\textsc{L}}^{\textsc{DP}} \geq \Opt_{\textsc{L}} - \alpha \,$ where $\alpha = \frac{4(u_{\max}-u_{\min})\sqrt{8n\log(\frac{n}{\delta})}}{\epsilon}$. 
\end{thm}
The proof of Theorem~\ref{thm} can be consulted in~\cite{hsu2014privately}.

The privacy cost in this study is defined in terms of the relative difference between $\Opt_{\textsc{L}}$ and $\Opt^{\textsc{DP}}_{\textsc{L}}$. In a sense, it exemplifies the diminished objective value arising as a result of solving the DR optimization problem with inaccurate customers demands obfuscated by differentially private noise. For {\sc UMDR$_\textsc{L}$}, the cost of privacy, denoted by $\Phi_\textsc{L} \in [0, 1]$, is defined as
\begin{align}
\Phi_\textsc{L} \triangleq \frac{\Opt_{\textsc{L}} - \Opt_\textsc{L}^{\textsc{DP}}}{\Opt_{\textsc{L}}}\,.
\label{phi}
\end{align}
When $\Phi_\textsc{L}=0$, this represents the ideal desirable case when no cost is incurred on the optimality of the produced energy management decisions. The closer $\Phi_\textsc{L}$ to $1$ the higher the privacy cost is and so is the optimality gap. For example, $\Phi_\textsc{L}=0.4$ implies that incorporating the privacy-preserving method entails $40$\% loss of optimality. Following the logic of Eqn.~\raf{phi}, define $\Phi$ to be the privacy cost for {\sc UMDR} problem.
\section{Empirical Evaluation}\label{results}

To complement the analytic result in Theorem~\ref{thm}, this section evaluates the utilized differentially private mechanism empirically by applying it to a simulated $4$-bus feeder from Canadian benchmark distribution system, which appears in~Fig.\ref{fig:system}. The objective is to investigate and quantify the privacy cost in the proposed event-based DR system. The {\sc Cplex} optimizer is utilized to obtain the close-to-optimal solutions numerically for {\sc UMDR}, {\sc UMDR$_\textsc{L}$} problems and their differentially private analogs.

\subsection{Simulation Setup and Settings}

The feeder, which is rated at $8.7$MVA, $400$A and $12.47$KV, is simulated with an overall generation capacity of $4$MVA and over $1500$ customers allocated among the nodes. Each feeder section is a $700$MCM Cu XLPE cable with impedance $z=0.1529 + J 0.1406$ $\Omega/km$. Each customer has a specific power demand (including both active and reactive power) and a utility that is generated according to a probability preference model. Due to limited power supply, the customers may suffer from a reduction of generation capacity occasionally. Various types of loads are considered including \textit{residential} and \textit{commercial} customers. Typically, the load power factor varies between $0.8$ to $1$ (to comply with IEEE standards) and thus the maximum phase angle between any pair of demands is restricted to be in the range of $[0, 36^{\circ}]$.

\subsection{Case Studies}
Various case studies are performed to evaluate the studied privacy-preserving mechanism considering diverse scenarios with respect to the correlation between customer loads and utilities, consumption profile and privacy levels. The following are settings for the case studies in this paper.

\begin{enumerate}
	\setlength{\itemindent}{0em}
	\item[(i)] {\em Utility-demand correlation}:
	\begin{enumerate}
		\setlength{\itemindent}{-1.3em}
		\item {\em Quadratic utility (Q)}: The utility of a customer is a quadratic function of the power demand in the form of $u_{k}^t({|S_k|}) = a\cdot {|S_k|}^2 + b\cdot {|S_k|} + c \,,~~\forall t \in \cT$, where $a > 0, b, c \ge 0$ are predetermined constants.
		\item {\em Uncorrelated setting (U)}: The utility of each customer is independent of the power demand and is generated randomly from $[0, |S_{\max}(k)|]$, where $|S_{\max}(k)|$ depends on the customer type.
	\end{enumerate}
	\item[(ii)] {\em Customer types}:
	\begin{enumerate}
		\setlength{\itemindent}{-1.3em}
		\item {\em Residential (R) customers}: The customer set is comprised of residential customers having small power demands ranging from $1500$VA to $15$KVA.
		\item {\em Mixed (M) customers}: The customer set is comprised of a mix of commercial and residential customers. Commercial customers have big power demands ranging from $300$KVA up to $1$MVA and constitute no more than $10$\% of all customers chosen at random.
	\end{enumerate}
	\item[(ii)] {\em Privacy levels}:
	\begin{enumerate}
		\setlength{\itemindent}{-1.3em}
		\item {\em Fixed (F) privacy level}: Customers are assigned only a single homogeneous level of privacy predetermined by LSE.
		\item {\em Variable (V) privacy level}: Customers are allowed to choose the desired privacy protection level from a predefined set offered by LSE.
	\end{enumerate}
	
\end{enumerate}

In this paper, the case studies will be represented by the aforementioned acronyms. For example, the case study named QMV stands for the one with mixed customers, quadratic utility-demand correlation and heterogeneous privacy levels.

\subsection{Simulations with Homogeneous Privacy Levels}

In this subsection the privacy cost is evaluated for case studies with homogeneous privacy levels, where each case study is analyzed considering changes in the set of customers. For each DR optimization problem (i.e., {\sc UMDR} and {\sc UMDR$_\textsc{L}$}), two different privacy levels are examined where the corresponding value of $\epsilon$ is set to $0.01$ and $1$. As for additive privacy parameter $\delta$, the value is fixed to $0.5$ for all case studies.  

\begin{figure}[!htb]
	\begin{center}
		\includegraphics[scale=.475]{{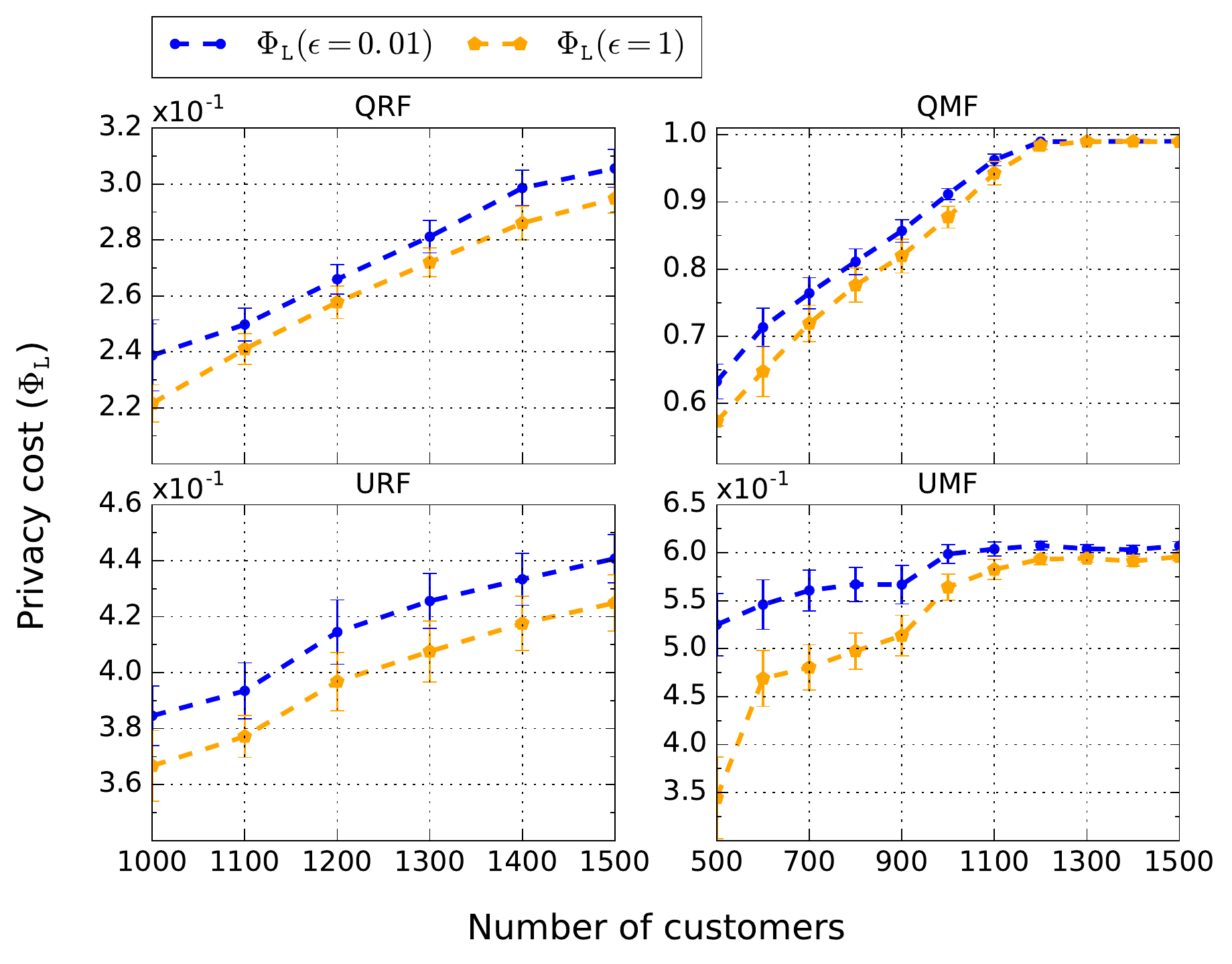}}
	\end{center}\vspace{-10pt}
	\caption{The average privacy cost for {\sc UMDR$_\textsc{L}$} against the number of customers at 95\% confidence interval.}
	\label{fig:1}
\end{figure}

The privacy cost for {\sc UMDR} is compared for different values of $\epsilon$ in Fig.\ref{fig:1}. The employed privacy method with an input parameter of is applied to the {\sc UMDR} problem (i.e., this corresponds to solving {\sc UMPDR} with perturbed customer utilities calculated accordingly) $30$ times for each of the $m$ number of customers, where $m$ varies between $500$ to $1500$ in steps of hundred. For case studies with residential customers, $m$ varies between $1000$ to $1500$ since with fewer customers the total load on MG is less than the supply thus resulting in trivial DR decisions. Each of $30$ iterations yields random changes in demands and utilities of customers.

%for case study QR (i.e., quadratic, residential) 

%It is worthy to note that the smaller these values are, the higher the privacy level guarantee.
%The optimal solutions computed by {\sc Cplex} optimizer for {\sc UMDR} and {\sc UMDR$_\textsc{L}$} problems, which correspond to non-private DR management, are considered to be the base case for the comparison. 

\begin{figure}[!htb]
	\begin{center}
		\includegraphics[scale=.49]{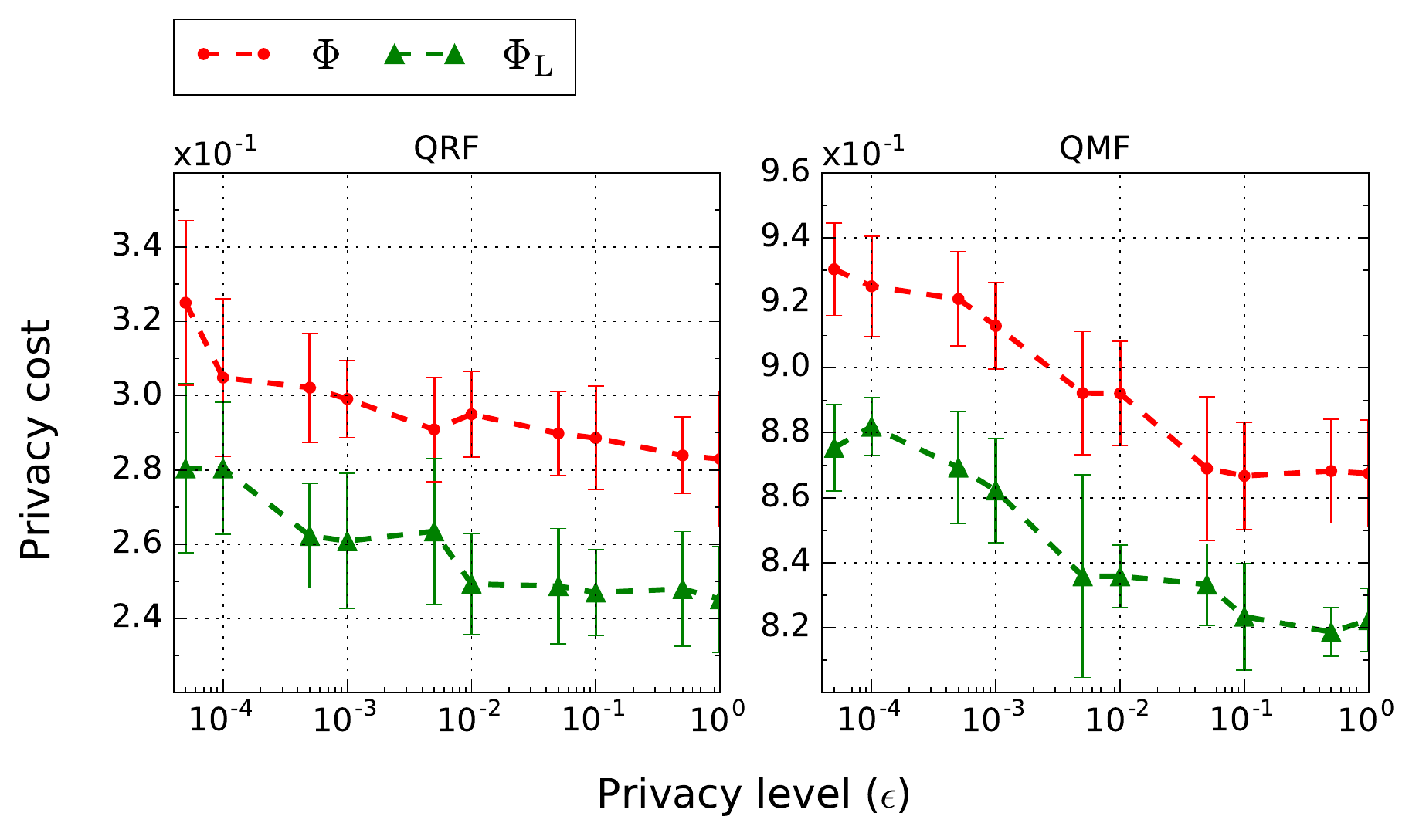}
	\end{center}\vspace{-10pt}
	\caption{The average privacy cost for {\sc UMDR} and {\sc UMDR}$_\textsc{L}$ against $\epsilon$ at 95\% confidence interval.}
	\label{fig:2}
\end{figure}

The privacy cost for {\sc UMDR} is contrasted with that of {\sc UMDR}$_\textsc{L}$ in Fig.~\ref{fig:2} considering a fixed number of customers. Here, the privacy level $\epsilon$ is varied from $5\cdot10^{-5}$ to $1$ for $900$ customers considering random changes in demands and utilities.  

One of the important findings observed is a common trend appearing in all the case studies performed. As can be inferred from Figures~\ref{fig:1} and~\ref{fig:2} increasing $\epsilon$ reduces the optimality gap and hence the privacy cost. In other terms, the higher the privacy requirements the higher the privacy cost. This directly implies that the privacy cost increases with the noise magnitude since the Laplace noise added to each customer's utility increases when lowering $\epsilon$. As illustrated in Fig.~\ref{fig:1}, on a MG with $500$ customers the studied privacy preserving mechanism resulted in a privacy cost of $0.59$ when $\epsilon=1$ (i.e., the optimality of DR is decreased by $59$\%), whereas when $\epsilon=0.1$ the privacy cost is about $0.63$, considering the case study QMF.

Another important observation is that with increasing customer participation optimality of produced DR solutions degrades drastically. As depicted in Fig.~\ref{fig:1}, when the number of customers is small the privacy cost is only in order of $0.6$ when considering case study QMF, while as the customer set cardinality grows privacy cost proliferates approaching nearly $1$. This highlights the necessity of devising efficient privacy preserving mechanisms, with a constant factor guarantee on the optimality gap, capable of solving large-scale DR management problems. 

Also, it is observed that the privacy cost is sensitive to different customer types. In particular, the privacy cost grows slower for case studies with residential customers than in those with mixed industrial and residential customers. As indicated in Fig.~\ref{fig:1}, the difference in privacy cost for case study QRF is about $0.08$ when considering the number of customers ranging from $1000$ to $1500$. In contrast, for the same range the privacy increases nearly by $1.2$ in case study QMF. In a sense, incorporation of privacy has more significant degrading effect on the optimality for the scenarios with mixed customers as compared to those with only residential customers.    

\subsection{Dynamic Generation Capacity and Heterogeneous Privacy Levels}
The previous set of experiments considered an MG with a fixed generation capacity and homogeneous privacy levels. Here simulations are performed considering the case when the
MG’s generation capacity is varying over time, which could be representative of a
case of hybrid system with renewable DG sources. Furthermore, in a real-world situation customers may desire different levels of privacy depending on their preferences. Thus, the simulation studies performed in this section consider heterogeneous privacy levels.

\begin{figure}[!htb]
	\begin{center}
		\includegraphics[scale=.46]{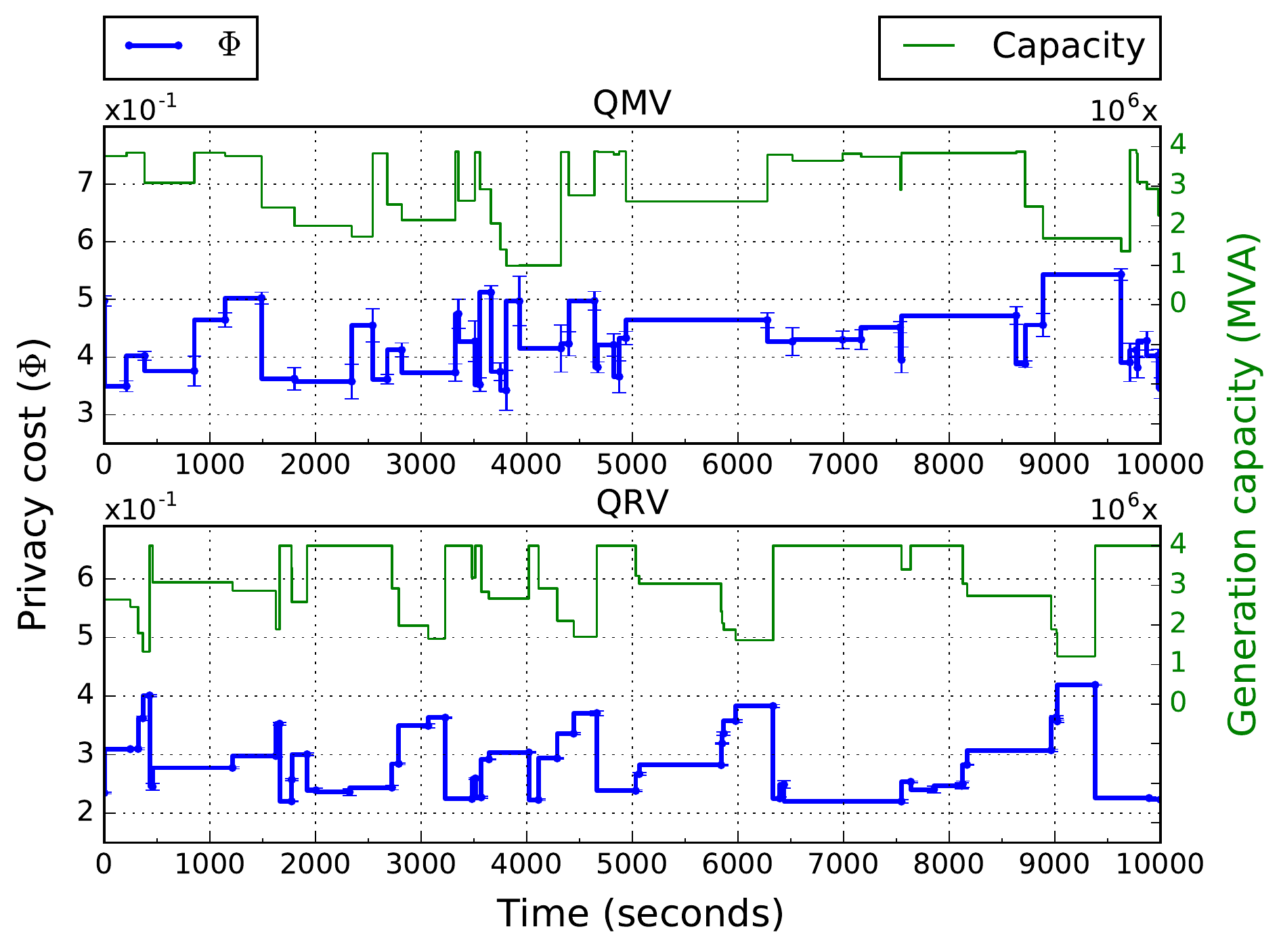}
	\end{center}\vspace{-10pt}
	\caption{The average privacy cost against the number of customers at 95\% confidence interval considering dynamic generation capacity.}
	\label{fig:time3}
\end{figure}

The time-varying generation capacity of MG that follows a Bernoulli process is dynamically varied between $1$MVA and $4$MVA from time $0$ to $10000$ (seconds). To analyze the effect of heterogeneous privacy levels, the employed mechanism is applied to the feeder with $1000$ customers. The observed privacy cost is plotted in Fig~\ref{fig:time3}.

\begin{figure}[!htb]
	\begin{center}
		\includegraphics[scale=.475]{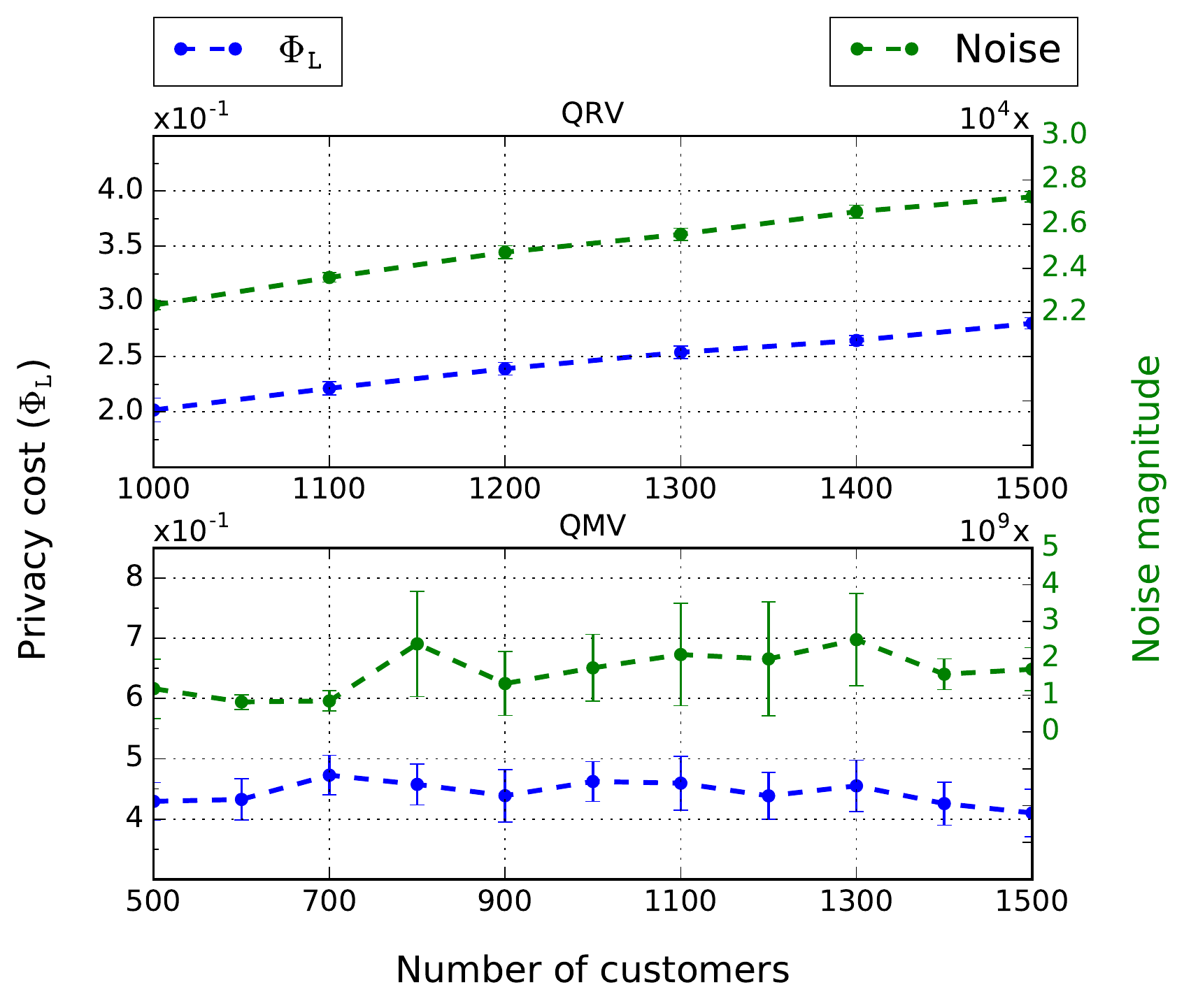}
	\end{center}\vspace{-10pt}
	\caption{The average privacy cost as a function of the number of customers at 95\% confidence interval.}
	\label{fig:time2}
\end{figure}

The privacy cost and absolute noise magnitude (per customer), which appear in Fig.~\ref{fig:time2}, are evaluated considering changes in the number of customers, whereas MG generation capacity remains fixed. 

The results demonstrate that allowing flexible privacy levels, instead of flat ones, may considerably increase the optimality of the proposed DR management scheme. The average privacy cost plotted in Fig.~\ref{fig:time2} for case study QMV fluctuates around $0.4$ while the number of customers changes and does not drift far away from that threshold. Whereas, in the scenarios with homogeneous privacy levels appearing in Fig.~\ref{fig:1}  the privacy cost was sometimes as high as nearly $1$.

\section{Conclusion}\label{concl}

This paper studies the trade-off between privacy and optimality in centralized DR management of MGs. Under the framework of differential privacy, the privacy cost is quantified empirically through extensive numerical simulations on a realistic distribution network. The observed results illustrate the striking effect posed on the optimality of produced solutions to DR optimization problem when considering increased privacy guarantees. According to the findings, the optimality gap approaches nearly $90\%$ in some cases, which urges the need for efficient privacy preserving algorithms with constant theoretical guarantees on the worst case performance. The major factors identified in this study that define this trade-off are the number of customers, privacy level and customer type. 

%The findings featured in this study shed light on the cost imposed on the quality of solution that arises with incorporation of privacy preserving mechanism into DR optimization problem. 
%\input{append}

\bibliographystyle{ieeetr}
\bibliography{reference}
\end{document}